\documentclass[%
 reprint,
superscriptaddress,
 amsmath,amssymb,
 aps,prl,
]{revtex4-1}

\usepackage{graphicx}
\usepackage{dcolumn}
\usepackage{bm}
\usepackage{amssymb, amsmath}
\usepackage{color}
\usepackage{amsfonts}
\usepackage{latexsym}
\usepackage{hyperref}

\newcommand{\unit}[1]{\ensuremath{\, \mathrm{#1}}}

\usepackage[english]{babel}
\usepackage[babel=true]{csquotes}

\def\eq#1{{Eq.(\ref{#1})}}    \def\fig#1{{Fig.\ref{#1}}}
\def\be{\begin{equation}}   \def\ee{\end{equation}}

\begin{document}


\title{Finite size effects in non-equilibrium membrane phase separation}

\author{Quentin Vagne}
\affiliation{Laboratoire Gulliver, UMR 7083 CNRS-ESPCI, 10 rue Vauquelin, 75231 Paris Cedex 05 - France.}
\email{quentin.vagne@espci.fr, pierre.sens@espci.fr}
\author{Matthew S. Turner}%
\affiliation{Dept. of Physics \& Complexity Centre, University of Warwick, Coventry CV4 7AL, UK.}
\author{Pierre Sens}
\affiliation{Laboratoire Gulliver, UMR 7083 CNRS-ESPCI, 10 rue Vauquelin, 75231 Paris Cedex 05 - France.}

\date{\today}

\begin{abstract}
The formation of dynamical clusters of proteins is ubiquitous in cellular membranes and is in part regulated by the recycling of membrane components. Mean-field models of out-of-equilibrium cluster formation with recycling predict a  broad cluster size distribution for infinite systems and must be corrected for finite-size effects for small systems such as cellular organelles. We show, using stochastic simulations and  analytic modelling, that tuning the system size is an efficient way to control the size of lateral membrane heterogeneities.  We apply these findings to a chain of enzymatic reaction sensitive to membrane protein clustering. The reaction efficiency is found to be a non-monotonic function of the system size, and can be optimal for sizes comparable to those of cellular organelles.
\end{abstract}


\maketitle

Cell membranes are heterogeneous structures in which different constituents undergo dynamical segregation~\cite{simons:1997,engelman:2005,lavi:2007}. The clustering of membrane components into dynamical domains is important for the control of enzymatic reactions necessary, e.g. for cell signalling~\cite{simons:2000b}. The thermodynamics of  composite membranes have been extensively studied {\em in-vitro}~\cite{bagatolli:2000,veatch:2002,baumgart:2003}. Below the critical point, line-tension driven phase separation results in the complete demixing of membrane components, after a coarsening process typically controlled by domain diffusion and coalescence. Cellular membranes on the other hand are out of equilibrium, and support continuous fluxes of matter. It has been shown that the     lateral size of cell membrane heterogeneities depends on the rate of membrane exchange, e.g.  exocytosis and endocytosis~\cite{tang:2001}. Mean field models of non-equilibrium phase separation with recycling predict a broad (power-law) distribution of domain size up to a maximum size directly controlled by the rate of membrane recycling~\cite{turner:2005}. For physiological recycling rates these models predict domain sizes that reach the typical size of cellular organelles such as endosomes or the Golgi apparatus, indicating the likely failure of the mean-field approach. In this letter, we study the effect of a finite system size on the steady-state size distribution of out-of-equilibrium membrane domains. Using an analytical approach combined with stochastic simulations we show that the system size is likely to crucially control the size distribution of membrane heterogeneities in cellular organelles and of membrane components that are corralled into regions of the plasma membrane. To illustrate the physiological relevance of our results, we show that the efficiency of a model of enzymatic reactions sensitive to enzyme clustering strongly depends on the  size of the compartment in which this reaction takes place, and that the optimal size can be comparable to the size of cellular organelles.

We first review the mean field solution of a diffusion and coalescence coarsening dynamics with external flux in an infinite system~\cite{turner:2005}. We consider a system populated by a minority species $A$ undergoing phase separation in a bulk of species $B$. The $A$ molecules diffuse with a diffusion coefficient $D$, aggregate whenever they meet to form domains of increasing sizes, which are recycled with some rate. The dynamics of coarsening by diffusion and coalescence has been extensively studied using the Smoluchowski coagulation equation \cite{smoluchowski:1916}, including in the presence of fluxes \cite{davies:1999,camacho:2001,connaughton:2004}. The influence of finite-size effects on the coalescence dynamics of membrane domains has recently been studied in a closed system {\it en route} to thermal equilibrium using a detailed hydrodynamic theory \cite{seki:2013}. However, the non-equilibrium steady-state of a finite-size system subjected to fluxes remains unexplored in spite of its central relevance to cellular organelles. Here we use a simplified version of the Smoluchowski  equation that neglects cluster fragmentation and the size dependence of the reaction rates, as is appropriate for 2D clusters with high line tension~\cite{turner:2005}. Areas are normalised by the area of a monomeric unit $s$, and time is expressed in unit of the typical diffusion time $\tau_{D}=s/D$:
\begin{eqnarray}
\frac{dc_{k}}{dt}=J_k-c_{k}\sum_{l=1}^{\infty}c_{l}+\frac{1}{2}\sum_{l=1}^{k-1}c_{l}c_{k-l}\cr
J_k=J(1-\phi)\delta_{k,1}-Kc_{k}
\label{mastereq}
\end{eqnarray}
Here, $c_k$ is the (dimensionless) concentration of clusters containing $k$ molecules of $A$, $\phi\equiv\sum_{1}^{\infty}kc_{k}$ is the surface fraction of $A$ components, and  $J_k$ is the source term. The latter consists of a constant influx of monomer  $J$ per unit area occupied by the $B$ species, and a uniform recycling of clusters of any size with a rate $K$. At steady state, the surface fraction converges to $\phi=J/(J+K)$, and the cluster size distribution, given in the Supplementary Information (SI), is well approximated by a power-law with an exponential cut-off~\cite{turner:2005}:
\begin{equation}
\label{approx}
c_{k}\approx \sqrt{\frac{\phi K}{2\pi}}k^{-\frac{3}{2}}\exp\left(-\frac{k}{\kappa}\right),\quad \kappa=\frac{2\phi}{K}
\end{equation}
The cutoff size $\kappa$ is completely determined by the recycling rates $J$ and $K$. One expects  this mean-field approach to break down whenever $\kappa$ is of order or larger than the system size.

\begin{figure}
\includegraphics[scale=0.4]{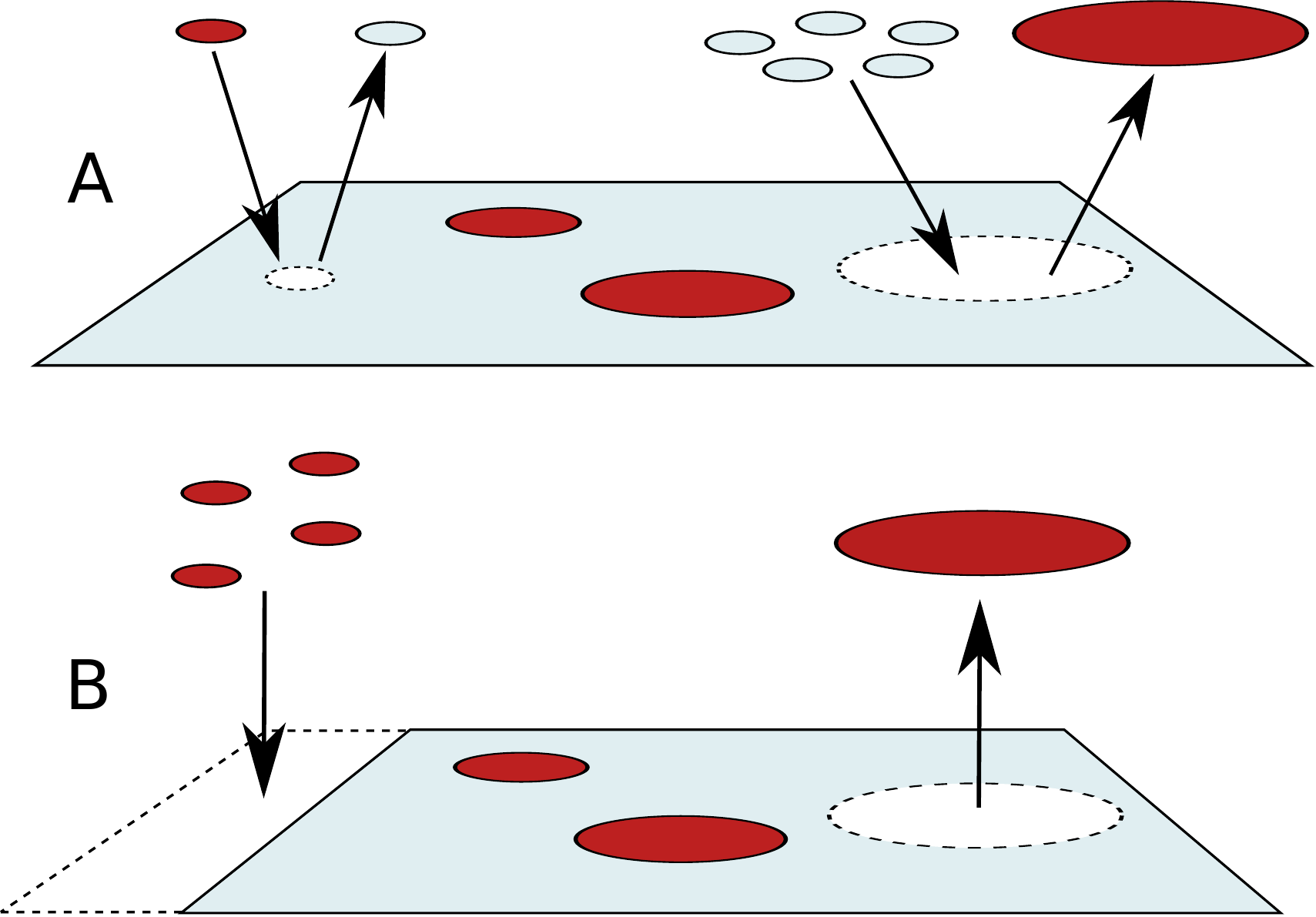}
\caption{\label{fig:budding} The two extreme models of recycling. In both cases monomeric units of $A$ components (red) are brought to the membrane while clusters of every size are recycled. (A) Fixed area model: The flux of $A$ is exactly counterbalanced at all times by an equal-and-opposite flux of $B$. (B) Fluctuating area model: $B$ components are not recycled, the area of $B$ is fixed and  the stochastic flux of $A$ components leads to fluctuations of the total membrane area.}
\end{figure}

The mechanisms by which cells regulate the size of their various organelles are complex and still poorly understood \cite{sengupta:2011,chan:2012,goehring:2012}. Here, we investigate two extreme scenarios of size regulation, see \fig{fig:budding}. In the {\bf Fixed area model} (\fig{fig:budding}A) there is perfect area regulation in which the flux of $A$ components is instantaneously compensated by an opposite flux of B components, thus keeping the total (dimensionless) membrane area $N_s$ constant. This is equivalent to a chemical reaction turning $B$ into $A$. Perfect regulation also requires that the extraction of a domain of size $k$ (denoted $A_k$) is compensated by the injection of $k$ components $B$. Under these assumptions the system dynamics are defined by the following rates:
\begin{equation}
\left\{
\begin{array}{r r c l}
\text{Injection} : & B & \xrightarrow{J} & A_{1} \\
\text{Growth} : & A_k + A_l & \xrightarrow{1/N_{s}} & A_{k+l} \\
\text{Removal} : & A_k & \xrightarrow{K} & k  B \\
\end{array}
\right.
\label{fixedtrans}
\end{equation}
At the mean field level, these processes correspond to the master equation \eq{mastereq}. In the alternative {\bf Fluctuating area model} (\fig{fig:budding}B) there is no such perfect regulation. Here the area of B component is constant and stochastic fluxes of $A$ result in fluctuations in the total membrane area. 
The (dimensionless) area covered by B components is $N_{s0}$, and the injection rate of $A$ monomers is assumed constant and independent of the system size. With $n_{k}$ the number of domains $A_{k}$, the system dynamics is defined by the following rates :
\begin{equation}
\left\{
\begin{array}{r r c l}
\text{Injection} : & B & \xrightarrow{J} & B+A_{1} \\
\text{Growth} : & A_k + A_l & \xrightarrow{1/N_{s}} & A_{k+l} \\
\text{Removal} : & A_k & \xrightarrow{K} & \emptyset \\
\end{array}
\right.
\label{flucttrans}
\end{equation}
The corresponding mean-field equation is slightly different from \eq{mastereq}, as the variations of the membrane size must be taken into account in the definition of the concentrations $c_{k}=n_{k}(t)/N_{s}(t)$. However, the resulting stationary state is the same as in the fixed area case (see SI), and is well approximated by \eq{approx}. 

{\bf Stochastic simulations}:
We implement a full stochastic version of the non-equilibrium clustering process using a Gillespie algorithms \cite{gillespie:1977} in order to study the influence of correlations and stochastic fluctuations introduced by the finite system size on the cluster size distribution. The system state is given by the set of stochastic variables $\Omega(t)=\{n_{1}(t),n_{2}(t),...\}$. We consider a Markovian evolution in which $\Omega(t+dt)$ is related to $\Omega(t)$ by transition rates obtained from either \eq{fixedtrans} or \eq{flucttrans} for fixed or fluctuating area respectively. 

\begin{figure}
\includegraphics[scale=0.35]{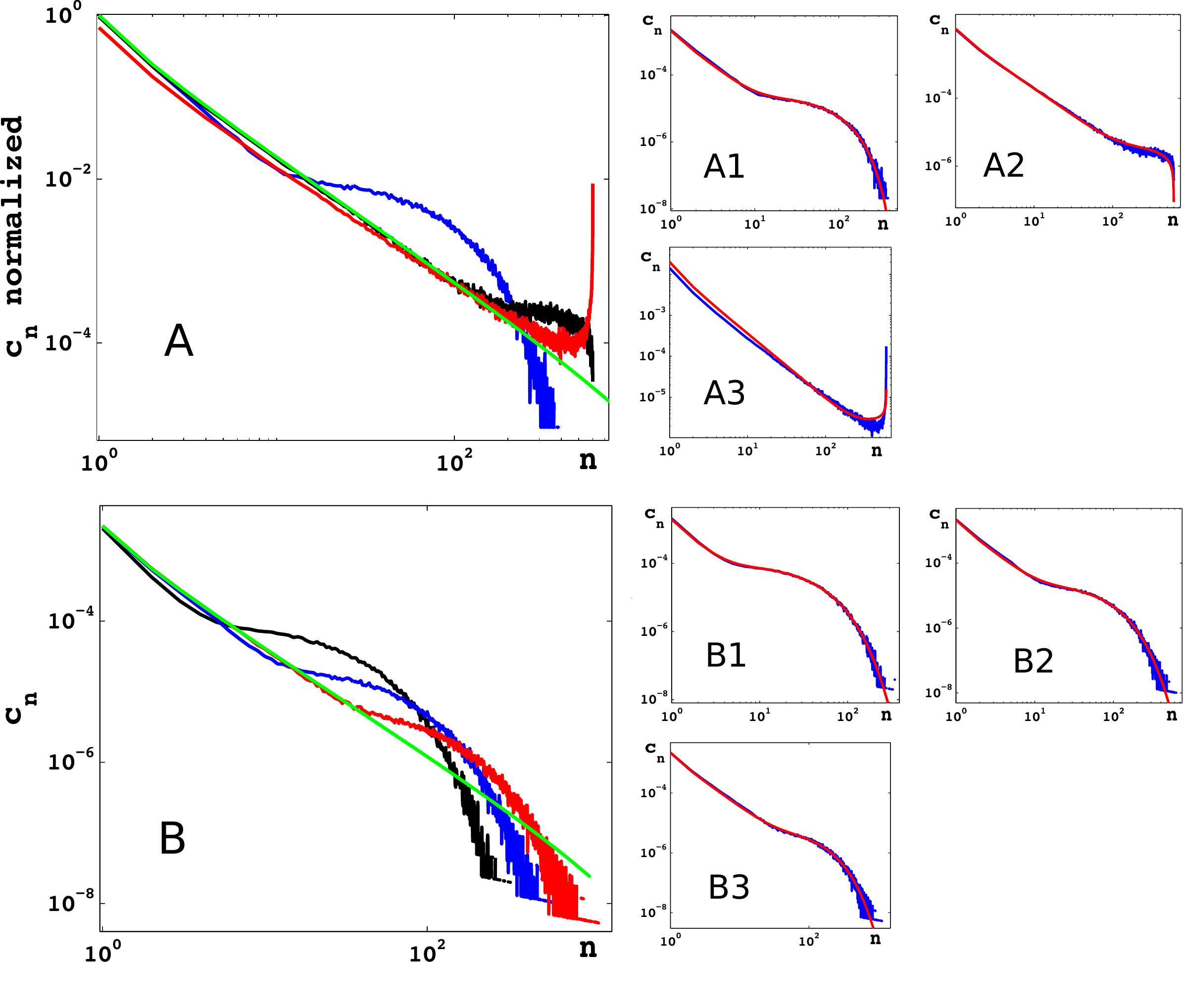}
\caption{\label{fig:deviations} Steady-state cluster size distributions obtained by simulations compared to the mean-field result (in green). (A) Results for the fixed area model ($N_{s}=600$, $\kappa=2000$), for the steady-state surface fraction $\phi$ equal to 0.1 (blue), 0.5 (black) and 0.9 (red). Panels A1-A3: Comparison of the simulation results (blue) with the analytical predictions of \eq{final} (red) for the three surface fractions. (B) Results for the fluctuating area model ($\phi=0.1$, $\kappa=2000$) with the initial membrane size $N_{s0}$ equal to 300 (black), 600 (blue) and 1200 (red). Results (not shown) at other surface fractions show no qualitative difference. Panels B1-B3: Comparison of the simulation results (blue) with the analytical predictions of \eq{final} (red) for the three initial sizes.}
\end{figure}

For membrane sizes large compared to the typical domain sizes ($N_s\gg \kappa$), the numerical results are in perfect agreement with the mean-field predictions at steady-state (see SI). However this is no longer the case when the membrane size is close to or smaller than the typical domain sizes, see \fig{fig:deviations}. The power-law predicted by the mean-field model \eq{approx} is still valid for small cluster sizes, but large deviations are observed beyond cluster sizes about one order of magnitude smaller than the system size. In the fluctuating area model, the features of the size distribution are similar for all surface fractions, and depends solely on the ratio $N_s/\kappa$. In the fixed area model the size distribution strongly depends on the steady-state surface fraction $\phi$.

\vspace{2mm}
{\bf Analytical model}: \fig{fig:deviations} shows average cluster concentrations. However, snapshots of stochastic simulations at a given time $t$ often show many small clusters and a single large one that grows until it gets recycled. This suggests  a mean-field approach to treat the population of small clusters and a stochastic approach for the large one. Let us call $c_k'$ the concentrations of the small domains and $p(n,t)$ the probability of having a large cluster of size $n$ at time $t$. We define $\phi_S=\sum kc_k'$ the average surface fractions of the small domains and $\phi_L$ the average surface fraction of the large cluster.
The detailed resolution of this simplified model is given in the SI, and is summarised below. Small domains are described by \eq{mastereq}, modified both because an average fraction $\phi_L$ of the total area is occupied by the large domain and because the effective recycling rate of small clusters  includes their rate of coalescence with the large cluster, so $K$ is replaced by $K+1/N_s$. The stochastic evolution of the size $n(t)$ of the large cluster is obtained using the following approximations; at each recycling event, $n$ is set to zero and a new \enquote{large} cluster is recreated from scratch by coalescence events, each of which is approximated to produce a unit step increase of $n$ (the large cluster being most of the time much larger than any small cluster). These assumptions lead to the stochastic evolution: $n\rightarrow n+1$ at a rate $\phi_S$ and $n\rightarrow0$ at a rate $K$.
The important difference between the fixed area and fluctuating area models is that in the former case, the remaining space populated by the small domains depends on the size $n$ of the large domain, which means that the growth rate of the large cluster depends on $n$. This leads to qualitatively different size distributions:
\begin{eqnarray}
&p_{\rm fixed}(n)\approx \frac{1}{N_s}\left(\frac{1-\phi}{\phi}+\frac{2N_s}{\kappa}\right)\left(1-\frac{n}{N_s}\right)^{\frac{1-2\phi}{\phi}+\frac{2N_s}{\kappa}}\cr
&p_{\rm fluct}(n)=\frac{K}{\phi_{S}}\left(\frac{\phi_{S}}{\phi_{S}+K}\right)^{n+1}
\label{pn}
\end{eqnarray}
the fixed area expression being  valid for $N_s\gg n\gg 1$.

The average surface fractions $\phi_S$ occupied by the small clusters is shown in the SI to be:
\be
\phi_S^{\rm fixed}=\frac{JKN_{s}}{(J+K)(KN_{s}+1)}\quad \phi_S^{\rm fluct} =\frac{JN_{s0}}{1+N_{s0}(J+K)}
\label{phis}
\ee
Using Eq.(\ref{approx},\ref{pn},\ref{phis}), the full size distribution is obtained as:
\be
c_{n,{\rm fixed}}=c_n'+\frac{p(n)}{N_{s}}\quad
c_{n,{\rm fluct}}=c_n'+\frac{p(n)(1-\phi_{S})}{N_{s0}+n}
\label{final}
\ee

The right side of Fig. \ref{fig:deviations} shows the comparison between the analytical expressions of \eq{final} and the simulation results. The analytical treatment captures the features of the two regulations schemes with excellent quantitative accuracy: a power-law distribution with a hump at large size for the fluctuating area model, and a similar shape at low $\phi$, but a non-monotonous distribution with a high probability of large cluster sizes for $\phi>1/(2(1-N_s/\kappa)$ for the fixed area model. We note that \eq{mastereq}, which treated the $A$ species as the minority phase, is not physically valid in the limit of high surface fraction.

\begin{figure}
\includegraphics[scale=0.4]{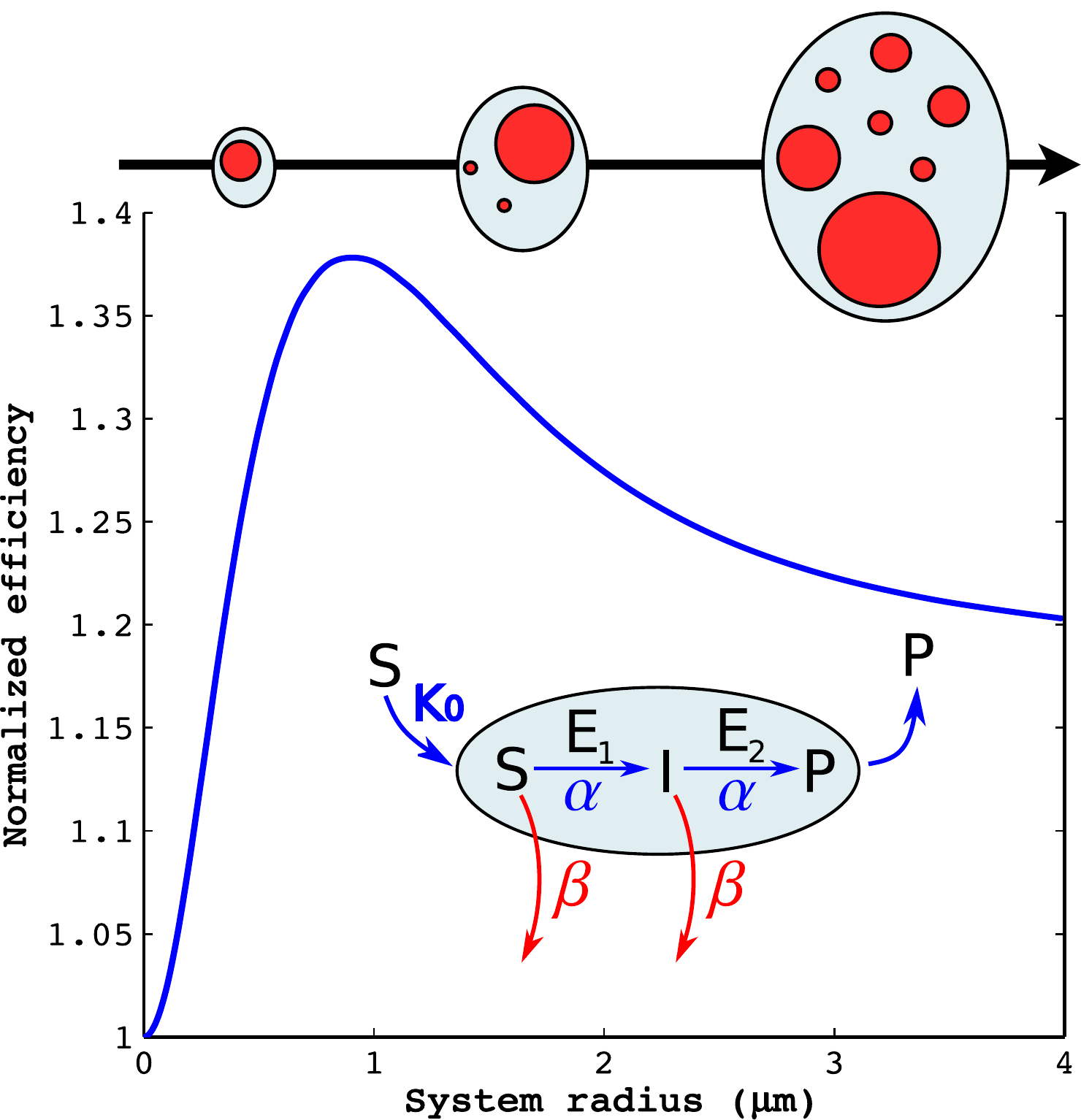}
\caption{\label{efficiency}Efficiency of a two-step enzymatic reaction taking place on the membrane of an organelle as a function of the organelle's size. The reaction is catalysed by two enzymes $E_1$ and $E_2$, both confined inside membrane domains with a system's size-dependent distribution shown in \fig{fig:deviations} (lower inset) and the trend for the size of the domain and the clusters therein are shown along the top. The efficiency $\eta^*$ (\eq{eq:eta}) is normalized by the efficiency of an homogeneous system $\eta_0$ \eq{etalimit}. All parameters are given in the text.}
\end{figure}

\vspace{2mm}
{\bf Application to enzymatic reactions}:
\fig{fig:deviations} shows that the size distribution of clusters of membrane components is strongly affected by finite size effects. 
In order to illustrate the  biological relevance of this result, we consider a sequence of enzymatic reactions taking place inside membrane domains. Biochemical transformations often involve several steps, catalysed by specific enzymes. The co-localization of different enzymes into clusters can improve reaction efficiency \cite{buchner:2013}. Moreover, it was recently shown that maximum efficiency can occur at an optimal cluster size \cite{castellana:2014}. Our results therefore suggest that the system size (e.g. the size of an organelle in the membrane of which a reaction takes place) could affect the efficiency of biochemical reactions requiring enzyme clustering. To explore this we consider the two-step reaction shown in the inset to \fig{efficiency} : a substrate $S$ is transformed by enzyme $E_{1}$ into an intermediate $I$  that is then transformed by enzyme $E_{2}$ into the product $P$. Both enzymes are assumed to be confined within membrane domains, and the two reaction rates, $\alpha$, are assumed to be identical for simplicity. The substrate is brought to the membrane homogeneously at a rate $K_{0}$ per unit area. We further assume that both the substrate $S$ and the intermediate $I$ are degraded (removed from the membrane) at the same rate $\beta$. 

The efficiency of the reaction $\eta$ is defined as  the ratio of the rate of production of $P$ over the rate of injection of $S$.
Defining the ratio of reaction to degradation rates $\mu\equiv \alpha/\beta$, it can be shown (see SI) that the efficiency for an homogeneous system $\eta_0$, and for an infinite system containing a single (infinite) domain $\eta_\infty$  are:
\be
\eta_0=\left(\frac{\phi \mu}{1+\phi \mu}\right)^2\quad,\quad \eta_\infty=\frac{\phi \mu^2}{(1+\mu)^2}
\label{etalimit}
\ee
Enzyme co-clustering into domains is beneficial because it increases the likelihood that the intermediate $I$, produced inside a cluster by an enzyme $E_1$, meets an enzyme $E_2$ before it gets degraded. On the other hand, clustering leaves large membrane patches free of enzymes, which increases the likelihood that the substrate $S$ is degraded before it meets an enzyme $E_1$. \eq{etalimit} shows that the balance between these competing effects favors clustering if  $\phi \mu^2<1$. This can be understood by noting that when $\mu$ and $\phi$ are very small most substrate and intermediate molecules are degraded before encountering an enzyme. Clustering of enzymes therefore increases efficiency.
 
The efficiency for intermediate cluster sizes is derived by dividing the system into subunits of size $R_c$ containing one domain of size $R_d\simeq R_c\sqrt{\phi}$ surrounded by enzyme-free membrane. We assume cyclindrical symmetry, and that the molecules $S$ and $I$ diffuse with coefficient $D_{m}\gg D$  so that domain diffusion is slow compared to the dynamics of $S$ and $I$. The concentrations $c_{S}(r)$ and $c_{I}(r)$ of $S$ and $I$ are given by
\begin{eqnarray}
\displaystyle \frac{dc_{S}}{dt} & = & K_{0}-\beta c_{S}-c_{S}\alpha\Theta(R_d-r)+D_{m}\Delta c_{S} \nonumber \\
\displaystyle \frac{dc_{I}}{dt} & = & -\beta c_{I}+(c_{S}-c_I)\alpha\Theta(R_d-r)+D_{m}\Delta c_{I}  \label{eq:diffreac}
\end{eqnarray}
with $\Theta$ the Heaviside step function. The efficiency of the subunit, defined by
\begin{equation}
\eta(R_d,\phi)=\frac{\alpha\int_{r<R_{d}}d\vec{r}c_{I}(\vec{r})}{K_{0}\pi {R_c}^{2}}\quad,\quad R_c=\frac{R_d}{\sqrt{\phi}}
\label{eta}
\end{equation}
is derived in the SI. The asymptotic efficiencies given \eq{etalimit} are recovered in the limit of very small and very large (dimensionless) domain size $R_{d}\sqrt{\beta/D_{m}}$. Importantly, the efficiency is maximised for a finite domain size provided $\mu>5/4$ and $\phi \mu <1$ (see SI). The larger $\mu$, the higher the gain in efficiency at the optimal domain size.

We compute the efficiency of a broad domain size distribution  $c_k$  using the  mean-field approximation (see SI):
\begin{equation}
\eta^*(N_s,\phi)=\frac{1}{\phi}\sum_{k=1}^{N_{s}}kc_k\eta(k,\phi)\quad , \quad\phi=\sum_{k=1}^{N_s}kc_k
\label{eq:eta}
\end{equation}
where $\eta(k,\phi)$ is given by \eq{eta} with $k\sim R_d^2$. The variation of the efficiency with the system size must be evaluated numerically. It is shown on \fig{efficiency} for a particular set of parameters. For the membrane parameters, we choose a unit size $s=1\unit{nm^{2}}$ \cite{nagle:2000} and a domain diffusion coefficient $D=0.1\unit{\mu m^{2}s^{-1}}$ \cite{cicuta:2007}, giving a typical diffusion time $\tau_{D}=s/D\simeq 10^{-5}\unit{s}$. Cellular organelles (the Golgi or endosomes) typically have their membrane fully renewed every few minutes \cite{wieland:1987}:  $K=0.1\unit{min^{-1}}$. Choosing $J=10^2\unit{\mu m^{-2}s^{-1}}$  gives a surface fraction $\phi\simeq 5\%$ and a domain cutoff size (for an infinite system) $\sqrt{s \kappa/\pi}\simeq 1.5\unit{\mu m}$. 
For the enzymatic reaction, we choose a diffusion coefficient $D_{m}=1\unit{\mu m^{2}s^{-1}}$, ten times larger than the domain diffusion coefficient $D$ in order to be in the range of validity of our approximation, but still in biological orders of magnitudes. We further choose $\beta=5\unit{s^{-1}}$ and $\alpha=25\unit{s^{-1}}$  which gives $\phi<1/\mu=0.2$. With this choice of parameter, the efficiency of a single unit (\eq{eta}) shows a maximum for intermediate-size domains $R_d=0.28\unit{\mu m}$. The global efficiency corresponding to the domain size distribution \eq{final} is shown in \fig{efficiency} as a function of the system size, and is maximal for a system radius of order $0.9\unit{\mu m}$, close to the typical size of cellular organelles, with a $38\%$ improvement compared to an homogeneous distribution of enzymes. 

The result shown in \fig{efficiency} illustrates how the size of a cellular compartment might regulate the efficiency at which a particular function is performed. Within our model the value of the efficiency depends on the clustering of membrane components, the size distribution of which is affected by the system size. Although we have chosen a particular sequence of enzymatic reactions, and we have chosen specific (and physiological) parameters, our work provides a quantitative analysis of a phenomenon that has the potential to be very general: Cell subdivision into organelles is a hallmark of eukaryotic cells. One reason for this is  to maintain different biochemical environments within the same cell. Another, we argue, could be that organelle's size is tuned to achieve higher enzymatic efficiency.
\vspace{4mm}
\begin{acknowledgments}
One of us (MST) acknowledges funding from the UK EPSRC under grant EP/E501311/1.
\end{acknowledgments}


%

\end{document}